\documentclass[sigconf]{acmart}
\usepackage{natbib}
\usepackage[utf8]{inputenc}
\pdfoutput=1

\acmConference[SIGCSE '21]{SIGCSE '21: ACM Technical Symposium on Computer Science Education}{March 02--05, 2022}{Providence, RI}
\copyrightyear{2022}
\acmYear{2022}
\setcopyright{licensedusgovmixed}\acmConference[SIGCSE 2022]{Proceedings of the 53rd ACM Technical Symposium on Computer Science Education V. 1}{March 3--5, 2022}{Providence, RI, USA}
\acmBooktitle{Proceedings of the 53rd ACM Technical Symposium on Computer Science Education V. 1 (SIGCSE 2022), March 3--5, 2022, Providence, RI, USA}
\acmPrice{15.00}
\acmDOI{10.1145/3478431.3499377}
\acmISBN{978-1-4503-9070-5/22/03}
\begin{document}
\title{Designing a Dashboard for Student Teamwork Analysis}
\author{Niki Gitinabard}
\email{niki@allobee.com}
\affiliation{%
  \institution{Allobee Inc.}
  \country{US}
}
\author{Sarah Heckman}
\email{sarah_heckman@ncsu.edu}
\affiliation{%
  \institution{North Carolina State University}
  \country{US}
}
\author{Tiffany Barnes}
\email{tiffany.barnes@gmail.com}
\affiliation{%
  \institution{North Carolina State University}
  \country{US}
}
\author{Collin F. Lynch}
\email{cflynch@ncsu.edu}
\affiliation{%
  \institution{North Carolina State University}
  \country{US}
}

\begin{abstract}
    Classroom dashboards are designed to help instructors effectively orchestrate classrooms by providing summary statistics, activity tracking, and other information \cite{roschelle13}. Existing dashboards are generally specific to an LMS or platform and they generally summarize individual work, not group behaviors. However, CS courses typically involve constellations of tools and mix on- and offline collaboration. Thus, cross-platform monitoring of individuals and teams is important to develop a full picture of the class. In this work, we describe our work on Concert, a data integration platform that collects data about student activities from several sources such as Piazza, My Digital Hand, and GitHub and uses it to support classroom monitoring through analysis and visualizations. We discuss team visualizations that we have developed to support effective group management and to help instructors identify teams in need of intervention.
\end{abstract}



\maketitle

\section{Introduction}
Modern CS courses are typically \textit{blended} and involve suites of online tools including learning management systems (LMSs) such as Moodle, Canvas, or Blackboard, development platforms such as GitHub, automated build servers such as Jenkins, and online support platforms such as My Digital Hand or Piazza. Students often face difficulties working within and across these platforms and integrating what they do online with their in-person learning \cite{sheshadri18}. To support students in using these tools and to \emph{orchestrate} good learning instructors must monitor the students' work \cite{roschelle13}. Many existing LMSs include tools for monitoring student activities collected using heat maps (e.g. \cite{bueckle17}) or other visualizations (e.g. \cite{emmons17,mazza07,sandee20,vivian15}). The majority of these tools, however, are focused on a single platform, often the course LMS, and do not allow the instructors to observe activities across platforms.

Existing monitoring is also complicated by teamwork. Collaborative projects are central to many CS courses as they provide opportunities for peer instruction and mirror professional environments.  While individual activities still make up the majority of coursework, effective teamwork is essential to students' success. However, many CS students, particularly those in early courses, are new to the concept of teamwork and struggle with the coordination, communication, and sharing that underpin teamwork \cite{feichtner84}. In order to effectively manage student teams, instructors must be able to monitor their work, identify patterns that cut across teams, and account for structured collaboration practices such as POGIL \cite{simonson19pogil} that allocate different tasks to different individuals. They must be able to distinguish cases where work is being delegated properly from cases of free-riding, or where the team itself is unable to communicate. Individual student analysis is unsuitable in such cases.

Our goal in this research is to address both challenges by implementing an interactive dashboard that can be used to integrate student information across platforms, visualize student and team activities, filter groups by instructor-defined criteria, and support data-driven intervention. We followed a design-based process in which we carry out initial focus interviews with experienced instructors at our university.  These interviews focused on the processes used by the instructors for group work, interventions, and the criteria that they use when making interventions, and preliminary platform designs. Subsequent to these interviews we implemented a working system based upon their comments.  We then returned to the instructors for subsequent interviews and evaluation of the system.

In this paper we first discuss related work and the background we drew from it.  We then describe our design interviews and the lessons we took from them, before turning to the system itself. We then describe our followup evaluation of the system and conclude with general lessons for similar work. At each stage we discuss the details of our process and lay out potential guidance.

\vspace{-0.2cm}
\section{Related Work}

With teamwork becoming central in many undergraduate CS courses, analyzing and understanding student teamwork becomes increasingly important. Team projects can help students gain the experience that helps their future in the industry \cite{feichtner84, reid05}, but such projects and the collaborative work can be new to many students, adding more complexity into the team assignments and team interactions \cite{feichtner84}. Instructors may help students learn teamwork by supervising and occasionally making interventions in their team activities, but finding and targeting struggling teams can be challenging for many instructors, often because of large class sizes.

Several instructor visualization tools have been developed and released with the aim of helping instructors to observe and understand their students' study behaviors and analyze and orchestrate their behaviors \cite{yoo15}.  Many of these dashboards focus on tracking student effort based upon their activities on a single online platform \cite{verbert13, sandee20,vivian15}. GitCanary is an example of such tools, monitoring project progress and contribution using activities on Git repositories \cite{sandee20}. Sandee and Aivaloglou provide a report of the iterative development of this tool based on student and instructor reports and an experimental evaluation of it during spring semester of 2020 in a class with 147 computer science students, guided by 7 teachers, working in teams of an average of four members, and developing an Android game. At the end of the semester they summarized teacher and student perceptions of the tool using six online interviews with the teachers and an anonymous questionnaire for the students. The results showed positive perceptions from both teachers and students, the tool being mainly used by teachers and students together for explaining concepts of software quality and to promote balance and planning. Vivian et al. also created a dashboard to extract team role distributions and emotions using term frequency in team discussions \cite{vivian15}. They tested this platform on Piazza data and were able to observe several different roles such as \textit{Communication}, \textit{Coordination}, \textit{Monitoring}, and \textit{Leadership}. Another popular platform is the Interactive Heat Map Analytics Dashboard which operates on the Canvas platform \cite{bueckle17}. This dashboard provides a multi-level heat map of individual student activities and performance as well as aggregate statistics.

Holstein et al. developed a similar dashboard called Luna, which utilized interaction data from intelligent tutoring systems used in a middle school \cite{holstein10}. They conducted a case study with 5 middle school teachers and 17 classes and showed that while the teachers often know how their students are doing, a learning dashboard based on an ITS data can improve their knowledge of the class \cite{xhakaj17}. They observed that the instructors primarily focused on information about the most challenging areas for the students. CourseVis is another example of graphical student monitoring tools, which works on a web-based platform for distance courses \cite{mazza07}. In this platform, they visualize social, behavioral, and cognitive activities of the students. The research group observed that using this platform, the instructors can identify trends in student activities and discover students who need help quickly. 

Some more recent dashboards have also incorporated different educational data mining and machine learning methods in the visualizations. Diana et al. for example, introduce a dashboard to provide instructors with real-time analytics about their programming assignments in Alice platform \cite{diana17}. In a follow-up work, they use the data collected from Alice and machine learning methods to predict students' performance and match low-performing students with high-performing peer-tutors \cite{diana18}. While Tarmazdi et al. designed a teamwork dashboard, which visualizes student roles \textit{Backup supporter}, \textit{Feedback provides}, or \textit{Leadership} in student teams \cite{tarmazdi15}. They use natural language processing methods to extract student roles from their online posts on the discussion forum. This work however, focused on the context of student discussions on the course forum and not on their coding activities.

While these dashboards provide valuable information about different aspects of the student activities, most are focused on individual student activities.  With the exception of Tarmazdi et al, the authors have not considered team-specific analytics. Moreover, these dashboards are focused on a single platform, typically a central LMS and do not integrate data from different platforms. While some LMSs such as Canvas and Moodle allow external tools to be linked in, they do not cover all tasks. CS courses typically use a collection of online tools, such as automated build interfaces, many of which do not support LMS integration.  In 2011, Siemens et al. proposed design of a platform for data integration and interventions \cite{siemens11}. They also proposed to develop distinct educator, learner, administrator, and researcher interfaces. However, due to lack of funding little progress was made towards implementing the platform \cite{muslim20}. The Concert platform that we develop in this work draws from multiple existing platforms including GitHub \cite{github}, Piazza \cite{piazza}, and My Digital Hand \cite{smith2017mdh} to track student actions across the course and we combine them to visualize the team activities for the course instructors. In future, we plan to extend the list of supported tools to include more online platforms used in classes.

In the current design of our system, we focused on student activities on Piazza and MDH (their discussion forum posts and office hour attendance record) as sources to understand team help-seeking patterns. These two sources reflect on student interactions inside teams and general interactions of the whole team with other teams or the teaching staff. We also looked into the amount of their activities on GitHub that reflects on how often they worked and how they divided the work among themselves.

\section{Methods}
Our goal in this work is to implement a platform to include data from different online sources, visualize student activities on them, and let instructors filter students based on their desired criteria. To do this, we scheduled two rounds of unstructured interviews with 9 instructors teaching at our university. The first round was done before the system implementation and the second round was scheduled after. During the first round of interviews, we asked the instructors about their criteria for making interventions in their classes and the kinds of interventions they make. We then shared some sketches of the system with them and asked what they would like to see included or changed in the platform implementation. During this discussion, the instructors shared some ideas to be included in the system, some concerns about possible challenges in implementation and decision making based on the included data, and some discussions about what data to include and what methods to use in the system. The main purpose of these interviews was to understand what patterns instructors generally look for in student activities, so that we can identify and present those patterns to them more easily.

In the following sections, we first discuss our interviews with these instructors about how they decide to make interventions in their classes and what kinds of interventions they make. The instructors were selected based on having experience with student teams in their classes. They teach a variety of courses such as introduction to computing, software engineering, programming concepts, senior design, game design, and HCI in the Computer Science department. We also talked to an instructor from the History department to gain insight about the tools and methods used outside computer science courses. Then, we discuss the design of the system and the features included based on the instructors' opinions. Finally, we share discussions from our second interviews with instructors where they interacted with the system and their suggestions for the future development of the platform.

\section{What Interventions Do Instructors Make in Their Classes?}

During the first round of interviews, we asked instructors if they ever plan interventions in their classes and what kinds of interventions they make. Two of the instructors stated that when they were concerned about student performance, they preferred to make interventions on a whole class basis, such as sending an email to the whole class encouraging them to be more active on Piazza or to come to office hours. One of these instructors mentioned that they prefer to remind students of different learning theories. For example: ``Theory says that the students who are more active on course discussion forums end up more successful in the class''. The reason for choosing top-level interventions was teaching style in one instructor's case and large size of classes in another. The second instructor mentioned that they have 60-70 teams in their class and it would be time-consuming for them to reach out to the teams individually. There were 6 other instructors that noted besides top-level interventions, they sometimes reach out to students or teams individually by email, set-up meetings with them, or refer them to their advisors if they are not having acceptable progress in course. Two of the instructors also mentioned that they sometimes make changes to the future course structure or the assignments based on the feedback they get from students. It is important to note that some instructors planned more than one kind of interventions in their classes. For example, one of the instructors reached out to individual teams and made changes to future courses. One of the instructors noted that they were not making any interventions in their class unless students reached out to them. 

We then asked what information they used to decide about when and to whom they reached out. For making top-level interventions, the instructors often considered low class attendance, low activity on the course forum, and low participation in assignments a red flag that often caused them to reach out to the whole class. For making changes to the future courses, the instructors noted checking out trends on the discussion forum and finding common concerns in class. For individual reach-outs, 6 instructors used student submitted forms such as peer evaluations, time-sheets, conflict reports, and evaluation forms; 4 instructors reached out if the students did not meet the course expectations regarding assignment submissions, project progress, or grades; and two instructors stated that they collect different statistics of the students and sometimes decide to make interventions based on those and their observations in class. This data included a summary of activities by project and student for one of them, and student grades as well as office hour and forum interactions for the other. Both these instructors noted that they do not have a formal process for making interventions and that their interventions are not always data-driven. 

After the first round of the interviews, we learned that interaction with individual teams in classes or using data-driven methods for finding struggling students and teams is time-consuming for most of the instructors. We also learned that the instructors considered a variety of factors for finding at-risk teams and the factors were even sometime different for the same instructor across different courses. As a result, we decided to keep the filters for selecting teams to contact and the time frames of data to include in the analysis a variable that the instructors can adjust based on their knowledge of their classes. We also decided to make it easy for them to check out student projects by including a link to each team's GitHub repository. Contacting teams was another part of the process that was time-consuming and thus we decided to make it easier by including email templates and the student emails for each project.

\section{System Design}

Our current platform is part of a larger ongoing project called \emph{Concert}. The Concert platform is a general data integration and intervention system that is designed to track students' work in courses across multiple platforms and to support rich models of their study habits, progress, and learning. This data is intended to support research on student learning, dashboards for instructor and student guidance, and data-driven interventions. The current version of the platform integrates data from Piazza (a discussion forum used widely in CS department of our university), Moodle (the LMS widely used for our courses), My Digital Hand (MDH, a ticketing system used in some large classes for keeping student turns in office hours) \cite{smith2017mdh}, and GitHub (a version control system used in some of CS courses for managing teamwork and submissions). For the present work we focused on data from the Software Development Fundamentals course (CSC 216) offered at NC State University. This course includes two multi-stage team projects in Java with intermediate milestones. We used data from the Fall 2020 iteration of the course for our work and we familiarized our participating faculty with the course design and expectations before the interviews.

We initially planned to filter the teams that we considered at-risk based on having large differences in activities with the course average and median values. However, there are many different patterns that can be identified across student teams in the studied course and using the data included in our system. These options will expand further once we include more courses and more sources of data in our system. So, it would be challenging to select some criteria and present teams meeting those to all the instructors. Some instructors might care about the overall progress of the project by the whole group, while others might get concerned about unbalanced work. Some courses might use GitHub for tracking student submissions and some might use other tools. So we decided to leave such decisions to the course instructors.

To keep the system analyses and team flagging measures dynamic, on the first page of the interface we ask instructors to select a course from their courses, select a time frame to filter activities by, and to select what aspects of student activities they would like to see. These can be subsets of discussion forum posts, office hour attendance, and submissions. We asked the instructors what kinds of date filters would work best based on their opinion. Some of our suggestions were week-by-week reports, all the activities since the start date of the project, and all the activities since the start of the semesters. Three of the instructors suggested that different courses might need different kinds of time frames and it is best to keep the time frames flexible for the users. For some courses with short project timelines, it might be best to include all the data since the beginning of the project. But some courses such as senior design have semester-long projects and it might be best if the user can see more recent activities. Thus, we decided to leave the time frame selection on the first page to allow the users to filter data based on their preferences.

\begin{figure}
    \centering
    \includegraphics[width=0.45\textwidth]{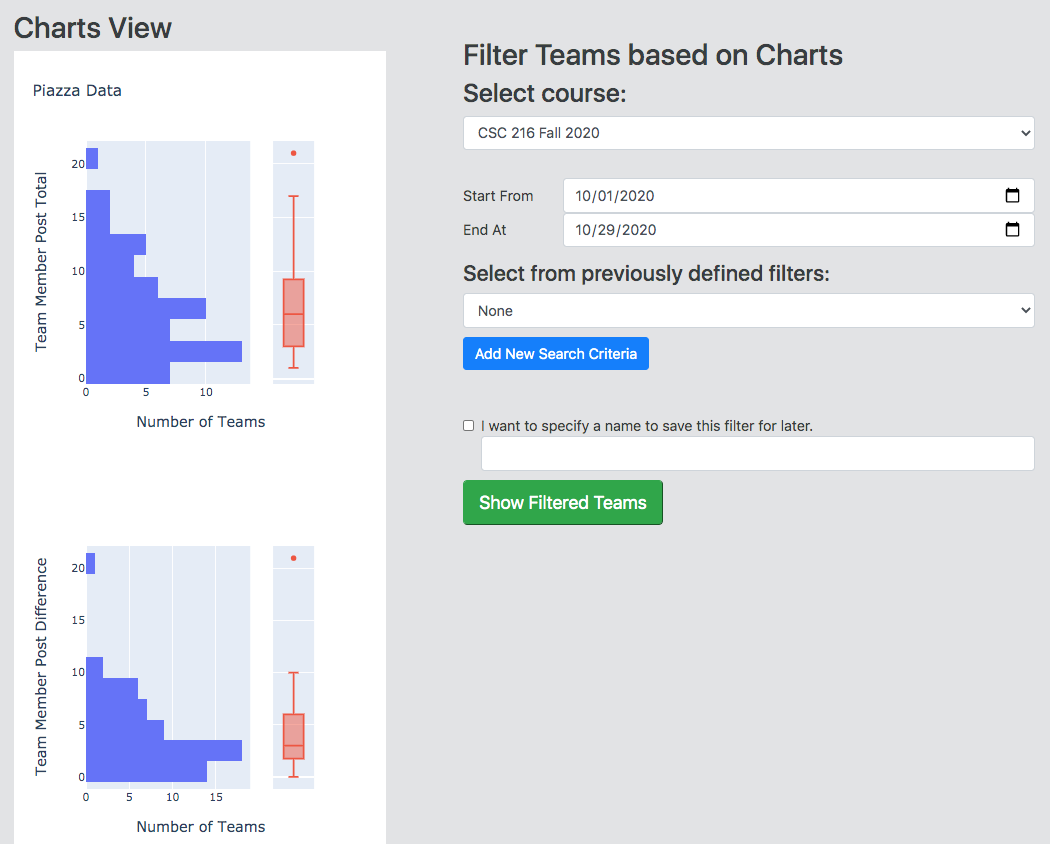}
    \caption{View of Team Charts}
    \label{fig:page2-1}
    \vspace{-0.45cm}
\end{figure}

\begin{figure}
    \centering
    \includegraphics[width=0.45\textwidth]{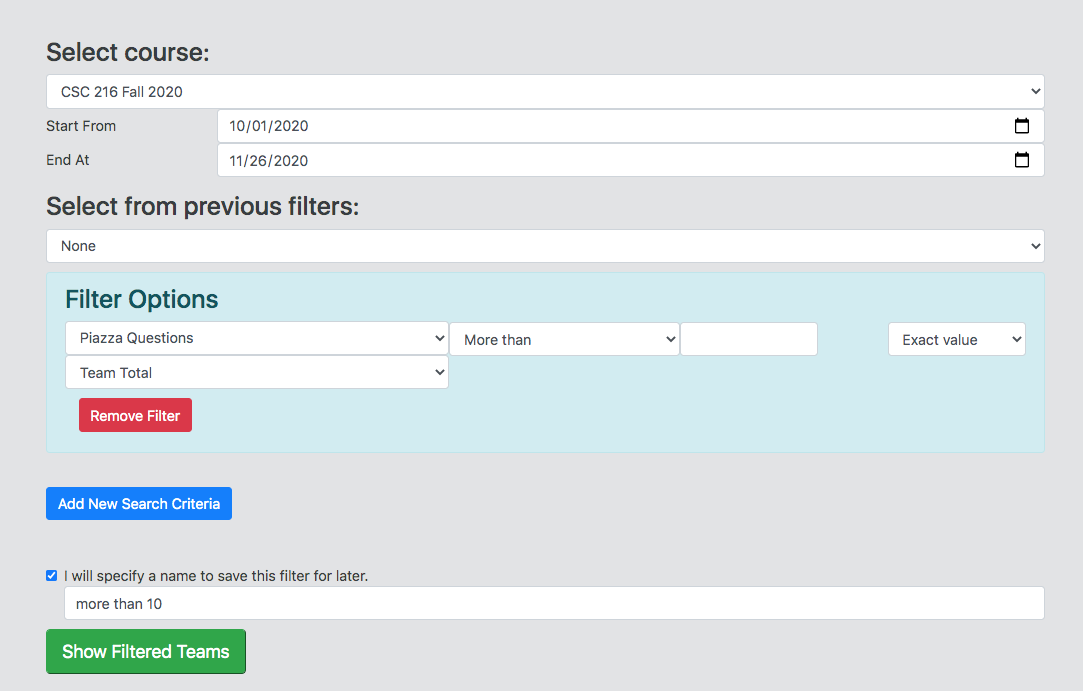}
    \caption{Filtering Page}
    \label{fig:page2-2}
    \vspace{-0.4cm}
\end{figure}

\begin{figure}[h]
    \centering
    \includegraphics[width=0.48\textwidth]{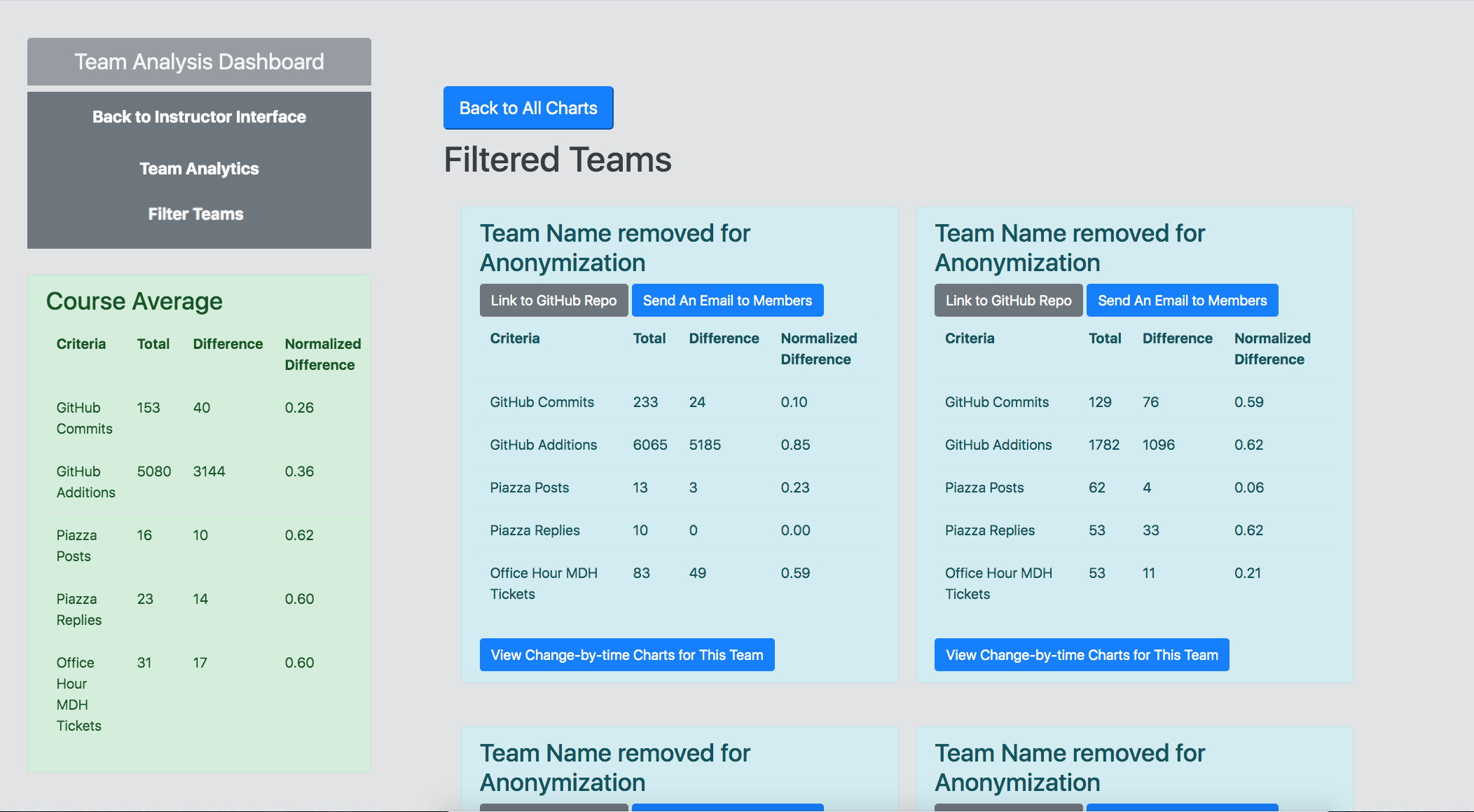}
    \caption{Overview of the Selected Teams}
    \label{fig:page3}
    \vspace{-0.4cm}
\end{figure}

On the next page after making these selections, we show some overall charts of the distributions of student activities for all the teams in class. We asked instructors if they would like to see total activities, absolute difference values between members, or normalized difference by the team activity and they noted that all of them together allow for more insightful decisions. Two instructors suggested that it would be best to differentiate between forum initial posts and replies. Three instructors also noted that it would be helpful to look at the amount of work done (by lines of code or Additions in commits) and not just the number of submissions (GitHub Commits). As a result, we have shown 5 categories of data on these charts: Piazza initial posts, Piazza replies, GitHub number of Commits, GitHub Additions or the lines of code added, and MDH tickets representing each time the team members asked a question in office hours. These charts were planned to help instructors identify patterns among team activities or points of concern such as large differences between team member submission sizes, general low progress of some teams, or low amounts of help-seeking activities in them. On the same page, we present filters based on the shown charts to choose student teams that the instructors might want to investigate further or reach out to. It is possible to define and mix several filters to include a combination of different metrics, for example finding teams that are not making much progress in their code but are also not asking for help on piazza or during office hours. An example of the charts shown on this page is included in Figure \ref{fig:page2-1} and an overview of the filtering page is shown in Figure \ref{fig:page2-2}. Once the instructors apply the filters, the list of the selected teams will appear on the next page, as shown in Figure \ref{fig:page3}. We also give instructors an option to choose a name and save the filter for later or use a previously saved filter as shown in Figures \ref{fig:page2-1} and \ref{fig:page2-2}.

\begin{figure}[h]
    \centering
    \includegraphics[width=0.48\textwidth]{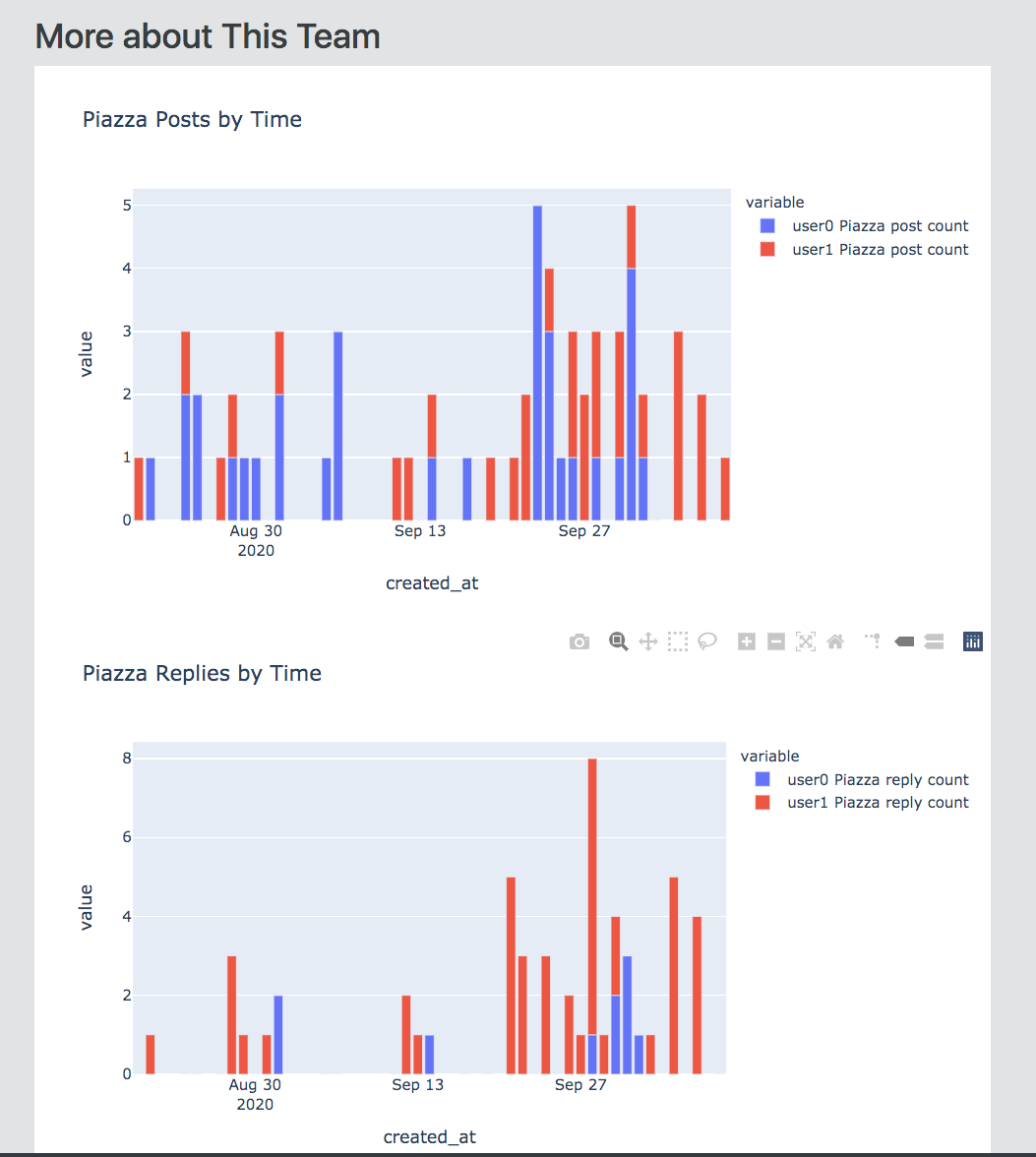}
    \caption{Team Details View}
    \label{fig:page4}
    \vspace{-0.4cm}
\end{figure}

For the selected teams, we show some general information such as their amount of work, their discussion forum posts and replies, and their office hour attendance. One of the instructors noted while using the system that it would be helpful to include course average values on this page so that it is easy for instructors to compare the selected teams to the course average. So, we included a table with course average values on the sidebar of this page. It was also suggested during the initial interview that a link to each student repository would be helpful in case the instructors decides to look at the team activities in more detail. Another suggestion made by instructors during the initial interviews was to prepare an email for the instructors to send to the team. This instructor noted that the users of the system will often have large number of students and teams in their classes and it will be time-consuming for them to compose and send emails to them one-by-one. This was in accordance with what another instructor noted about not being able to make individual interventions in their class due to the size of the class. Based on these suggestions we included a link to each team's repository on GitHub and also offered an option to send the team an email. Once the email button is clicked, an email draft will open on the user's page, including the team members' email addresses. We also include the email subject and content, modified by the names of the students in the team, which the instructors can edit before sending. Currently, the template for the emails is pre-defined in the system but we plan to extend this and  let the users define and save their templates, leaving place-holders for the system to fill. 
One of the instructors noted that showing timelines of activities for teams can also be helpful. To not make this page too crowded, we provided a link for each team to show more details of student activities in their team. An overview of the selected teams page and the team specific details page are shown in Figures \ref{fig:page3} and \ref{fig:page4} consecutively.

\section{Expected Interventions}

When we showed the design sketches of the system to the instructors, we asked them what kinds of interventions they see likely to make based on the information provided. The instructors who preferred top-level interventions stated that they still prefer to reach out to the whole class and educate them about effective teamwork. However, both of them as well as another instructor stated that they might reach out to individual teams if it is necessary. Two of the instructors believed unbalanced work is an issue that can be found by this platform and they would reach out to teams if they found unbalanced working among them. Three instructors noted that they would use filters to flag some teams and further observe and investigate them. Three instructors noted they might set up meetings with the students and ask questions based on their observations. They believed they can use the team-specific information shown on the platform as talking points during the meetings or maybe even share it with the students to express the reason for their concerns.  

One of the instructors was specifically interested in making changes for the future semesters. They stated that previously, they were able to observe role specification during the main course project when each member would work independently and not learn about other parts. This specialization caused some of the students to not learn about some aspects of the class. As a result of this observation, they added an individual project to that class to be done before the team project and believed that this could help different students to learn about different goals of the course. They stated that having this tool can help them identify such trends in their classes and consider solutions for them for the same semester or the future ones. They also mentioned that finding students who are active on the forum can help them identify possible peer tutors or future TAs.

Another instructor also mentioned that having this tool can help them change the interventions in their classes from ad-hoc to real-time. They noted that many interventions in their classes are based on student peer-evaluations and complaint forms, which are often submitted after the assignment is done. Being able to observe student activities real-time would help them identify such teams earlier and possibly provide them with help.

\section{System Evaluation and Use}
During the second interview, we presented the early implementation of the system to the instructors and recorded the kinds of activities they performed, the kinds of insights they could extract, what they liked most about the system, and their overall opinion of the system. We will discuss each of these subjects in this section. We also collected their suggestions for planning the future development of the project, which we will discuss in Section \ref{sec:future}.

\subsection{Instructor Queries and Insights}
The instructors applied different time-frames for observing student activity charts. These time-frames included windows as short as one week and as long as the whole semester. Searching for the data of the whole semester was the most popular as 5 out of 9 instructors searched at-least once for those.

Among different sources, the instructors again took different selection approaches. A few of them preferred to start with looking at all the available charts, while others limited the charts to only one kind at each search, looking for either discussion forum activities, office hour attendance, or work submissions.

They used the filters for finding many different trends among the student activities. Some examples of these trends are shown below:
\begin{itemize}
    \item Teams who are not asking for any kind of help by not going to office hours or asking questions on Piazza.
    \item Teams that had unbalanced work, who had large differences or normalized differences in team based on the number of commits and lines of code.
    \item Teams that are active on Piazza but not going to office hours.
    \item Teams that have only one member who is comfortable with Piazza.
    \item Teams that are most active on Piazza.
\end{itemize}

By applying these search measures, they were able to evaluate assumptions about the different teams. For example, the instructor who searched for teams that were active on Piazza but did not attend office hours, hypothesized that these teams were either too shy to seek in-person help or had scheduling conflicts. Also, by examining the detailed activities of the teams with larger differences, the instructors found that often one member did significantly more amount of work. They also stated that having the GitHub link would be useful for further investigation, since it is possible that one member submitted a large library or was fixing a bug which caused their large amount of activities. One of the instructors was interested in the members with large amounts of Piazza replies, stating they are the ones helping others most. They also noticed a team with 18 office hour visits and large number of GitHub activities, saying they seemed to be learning a lot. By looking into team details, they were also able to note the teams that had one member submitting more code and the other member attending more office hours, which looked like dividing the work. By looking at the normalized difference chart for all teams, one instructor noticed that it looks like there is one person who feels more comfortable on Piazza for most teams, since there was a large group of teams with a normalized difference close to 1. They also found some teams with similar number of commits and large amount of submitted lines of code and noted that one member might have submitted the project GUI which was provided by the instructors.

The instructors were also able to see specific trends in team activities based on their team specific charts, such as large spikes that the instructors guessed would be the course milestones and deadlines. The course instructor was able to confirm that those dates were in fact deadlines for different milestones of the projects. The two large projects of the course were visible by the trends of student activities during the whole semester.

\subsection{Instructor Assessments}
6 out of 9 instructors stated that the pre-filled emails were the best feature on the platform, especially if they would be able to prepare templates for them at the beginning of the semester. They mentioned that once you spot teams that might need interventions, sending emails to them would be the most time-consuming part and having this option would save them a lot of time.

One instructor liked the filtered team stats shown in tables. They stated that having information on one aspect of student activities is not as helpful as having the integrated information from all the different sources. They believed it would provide them with lots of information about the teams. Two instructors liked the single-team charts, since it showed them the trend and distribution of student activities in time. Two instructors noted that having the link to GitHub was specifically useful, since sometimes you need to check out their activities to make sure what the numbers mean. While numbers can draw your attention to some teams, you still sometimes need to investigate further.

Three instructors believed that most of the features were there and the system was already providing value. While they believed some design changes and providing more information about how to read different charts and statistics could make the system easier to start with or use during time, they mentioned that they got used to finding the information easily. Three other instructors stated that the system was helpful and it was useful to know how the students worked in teams. One of them stated that GitHub provides the users' pulse and you can see what each member did, but integrating that information with Piazza and office hours can be insightful. Two instructors stated that they would use the system if it would provide more charts and information at the starting page and without needing much time from them. Also, all the instructors believed that having the option of saving and re-using the filters makes the system more time-saving and easy to use.

\section{Conclusions and Future Work}
\label{sec:future}
In this work, we conducted detailed design interviews with 9 instructors about their courses, teamwork, and how they make data-guided interventions. Guided by this information we solicited their opinions on our proposed dashboard design and operation. We then used their feedback to guide our implementation of a working prototype using integrated data from a CS2 course. This dashboard provided the instructors with an overall view of their classes and provided them with tools to query data about specific teams or set thresholds to identify students who may need support. We then trained the instructors on the prototype and conducted additional design interviews to solicit their opinions on the platform. While providing this information to the instructors does not guarantee to improve student teamwork, it will save instructors the time they would need to spend analyzing student online behavior.

Overall, the instructors found the platform to be useful and believed that the information shown on this dashboard could help them to better identify and engage with struggling teams. Some of them mentioned that they would reach out to the top teams as well, to motivate them in keeping their good work going. The instructors also provided us with feedback about what other features they would like the system to have and what changes could make the system easier to interact with. Some of the feedback instructors provided was about the design of the pages and the information shown. For example, the charts did not seem self-sufficient without explanation. The axis labels and chart names were not descriptive enough and some of the defined metrics such as difference and normalized difference needed a more clear definition for the first-time users.

Some of the other suggestions were about the features we could add to the system to make it more useful and time-saving. The most asked for feature was being able to generate and use different email templates. The instructors believed it would save them more time and give them more flexibility to reach out to different groups of students, maybe just to say they are doing great. They also asked for the option to reach out only to the student who is less or more active in a team individually. Aside from saving and re-using filters, one instructor asked to set up filters at the beginning of the semester and get notifications about them regularly. Another instructor suggested that once filters were saved and given their names, we would be able to construct more complex filters by grouping them into logical expressions. One other feature the instructors requested more often was having pre-defined filters to show students above, within, and below median range of submissions so that they can find them without needing to create filters. Another request was having a landing page with insightful information about the class, such as general information, list of teams and how much activities they have done, or some charts providing information about the class at a glance.

There were also different requests for more personal configurations, such as the default time-frame setting or a landing page with useful insights about the course. Some instructors also asked for different kinds of visualizations and being able to switch between them. For example, two instructors preferred to see bar charts with a bar for each team instead of the distributions, to know where each team stands by looking at the chart. One other instructor also preferred to have the bar charts in team specific pages not stacked to make them easier to compare, but they believed it would be best if they could switch between stacked charts and bar charts. Some instructors asked for more filtering options such as getting keywords or names of the folders to limit the Piazza posts and office hour visits to the ones related to the team project. Another example was for the instructors to provide a regular expression for the files to ignore in submissions, since sometimes parts of the project code is provided by the teaching staff and its submission by one member can make the work distribution inaccurate. Instructors also requested for the project deadlines and milestones to be added to the system and be shown as options for time-frame selection and/or as overlays on the charts to give the user more insight about student activities.

One instructor suggested showing different types of office hour attendance with different colors on the chart. When students submit a request on My Digital Hand platform, there are three possibilities. They may be helped by the teaching staff present and have their problem resolved, they may be helped but leave without having their problem solved, or they may not get to be helped due to the office hours being busy. Showing these groups separately can help the instructors know if they have enough office hours and if the students are receiving the help they need.

We plan to implement different parts of these suggestions in the next phases of development, prioritizing the ones most asked for such as making the visualizations easier to follow, having predefined filters that are easy to use, and saving and selecting different email templates for reaching out to students more easily.

\begin{acks}
This research was supported by NSF \#1821475 ``Concert: Coordinating Educational Interactions for Student Engagement'' Collin F. Lynch, Tiffany Barnes, and Sarah Heckman (Co-PIs).
\end{acks}

\newpage
\bibliographystyle{abbrv}
\bibliography{SIGCSE}

\begin{thebibliography}{10}

\bibitem{github}
Github.
\newblock \url{https://github.com/}.
\newblock Accessed: 2021-08-11.

\bibitem{piazza}
Piazza discussion forum.
\newblock \url{https://piazza.com/}.
\newblock Accessed: 2021-08-11.

\bibitem{bueckle17}
M.~Bueckle and K.~B{\"o}rner.
\newblock Empowering instructors in learning management systems: Interactive
  heat map analytics dashboard.
\newblock {\em Retrieved Nov}, 2:2017, 2017.

\bibitem{diana17}
N.~Diana, M.~Eagle, J.~Stamper, S.~Grover, M.~Bienkowski, and S.~Basu.
\newblock An instructor dashboard for real-time analytics in interactive
  programming assignments.
\newblock In {\em Proceedings of the Seventh International Learning Analytics
  \& Knowledge Conference}, pages 272--279, 2017.

\bibitem{diana18}
N.~Diana, M.~Eagle, J.~Stamper, S.~Grover, M.~Bienkowski, and S.~Basu.
\newblock Peer tutor matching for introductory programming: Data-driven methods
  to enable new opportunities for help.
\newblock International Society of the Learning Sciences, Inc.[ISLS]., 2018.

\bibitem{emmons17}
S.~R. Emmons, R.~P. Light, and K.~B{\"o}rner.
\newblock Mooc visual analytics: Empowering students, teachers, researchers,
  and platform developers of massively open online courses.
\newblock {\em Journal of the Association for Information Science and
  Technology}, 68(10):2350--2363, 2017.

\bibitem{feichtner84}
S.~B. Feichtner and E.~A. Davis.
\newblock Why some groups fail: A survey of students' experiences with learning
  groups.
\newblock {\em Organizational Behavior Teaching Review}, 9(4):58--73, 1984.

\bibitem{holstein10}
K.~Holstein, F.~Xhakaj, V.~Aleven, and B.~McLaren.
\newblock Luna: a dashboard for teachers using intelligent tutoring systems.
\newblock {\em Education}, 60(1):159--171, 2010.

\bibitem{mazza07}
R.~Mazza and V.~Dimitrova.
\newblock Coursevis: A graphical student monitoring tool for supporting
  instructors in web-based distance courses.
\newblock {\em International Journal of Human-Computer Studies},
  65(2):125--139, 2007.

\bibitem{muslim20}
A.~Muslim, M.~A. Chatti, and M.~Guesmi.
\newblock Open learning analytics: a systematic literature review and future
  perspectives.
\newblock {\em Artificial Intelligence Supported Educational Technologies},
  pages 3--29, 2020.

\bibitem{reid05}
K.~L. Reid and G.~V. Wilson.
\newblock Learning by doing: introducing version control as a way to manage
  student assignments.
\newblock In {\em Acm Sigcse Bulletin}, volume~37, pages 272--276. ACM, 2005.

\bibitem{roschelle13}
J.~Roschelle, Y.~Dimitriadis, and U.~Hoppe.
\newblock Classroom orchestration: Synthesis.
\newblock {\em Computers \& Education}, 69:523--526, 2013.

\bibitem{sandee20}
J.~J. Sandee and E.~Aivaloglou.
\newblock Gitcanary: A tool for analyzing student contributions in group
  programming assignments.
\newblock In {\em Koli Calling'20: Proceedings of the 20th Koli Calling
  International Conference on Computing Education Research}, pages 1--2, 2020.

\bibitem{sheshadri18}
A.~Sheshadri, N.~Gitinabard, C.~F. Lynch, T.~Barnes, and S.~Heckman.
\newblock Predicting student performance based on online study habits: A study
  of blended courses.
\newblock In {\em Proceedings of the 11th International Conference on
  Educational Data Mining (EDM) 2018, Buffalo, US}, pages 411--417, 2018.

\bibitem{siemens11}
G.~Siemens, D.~Gasevic, C.~Haythornthwaite, S.~Dawson, S.~B. Shum, R.~Ferguson,
  E.~Duval, K.~Verbert, R.~Baker, et~al.
\newblock {\em Open Learning Analytics: an integrated \& modularized platform}.
\newblock PhD thesis, Open University Press Doctoral dissertation, 2011.

\bibitem{simonson19pogil}
S.~R. Simonson.
\newblock {\em POGIL: An introduction to process oriented guided inquiry
  learning for those who wish to empower learners}.
\newblock Stylus Publishing, LLC, 2019.

\bibitem{smith2017mdh}
A.~J. Smith, K.~E. Boyer, J.~Forbes, S.~Heckman, and K.~Mayer-Patel.
\newblock My digital hand: A tool for scaling up one-to-one peer teaching in
  support of computer science learning.
\newblock In {\em Proceedings of the 2017 ACM SIGCSE Technical Symposium on
  Computer Science Education}, pages 549--554, 2017.

\bibitem{tarmazdi15}
H.~Tarmazdi, R.~Vivian, C.~Szabo, K.~Falkner, and N.~Falkner.
\newblock Using learning analytics to visualise computer science teamwork.
\newblock In {\em Proceedings of the 2015 ACM Conference on Innovation and
  Technology in Computer Science Education}, ITiCSE '15, page 165–170, New
  York, NY, USA, 2015. Association for Computing Machinery.

\bibitem{verbert13}
K.~Verbert, E.~Duval, J.~Klerkx, S.~Govaerts, and J.~L. Santos.
\newblock Learning analytics dashboard applications.
\newblock {\em American Behavioral Scientist}, 57(10):1500--1509, 2013.

\bibitem{vivian15}
R.~Vivian, H.~Tarmazdi, K.~Falkner, N.~Falkner, and C.~Szabo.
\newblock The development of a dashboard tool for visualising online teamwork
  discussions.
\newblock In {\em 2015 IEEE/ACM 37th IEEE International Conference on Software
  Engineering}, volume~2, pages 380--388. IEEE, 2015.

\bibitem{xhakaj17}
F.~Xhakaj, V.~Aleven, and B.~M. McLaren.
\newblock Effects of a teacher dashboard for an intelligent tutoring system on
  teacher knowledge, lesson planning, lessons and student learning.
\newblock In {\'E}.~Lavou{\'e}, H.~Drachsler, K.~Verbert, J.~Broisin, and
  M.~P{\'e}rez-Sanagust{\'i}n, editors, {\em Data Driven Approaches in Digital
  Education}, pages 315--329, Cham, 2017. Springer International Publishing.

\bibitem{yoo15}
Y.~Yoo, H.~Lee, I.-H. Jo, and Y.~Park.
\newblock Educational dashboards for smart learning: Review of case studies.
\newblock {\em Emerging issues in smart learning}, pages 145--155, 2015.

\end{thebibliography}
\end{document}